# What elements should we focus when designing immersive virtual nature? A preliminary user study

Lin Ma[1], Qiyuan An[1], Jing Chen[1], Xinggang Hou[1], Yuan Feng[1], Dengkai Chen[1, *]

1. School of Mechanical Engineering, Northwestern Polytechnical University, 127 West Youyi Road, Beilin District, Xi'an Shaanxi, 710072, P.R.China.

## 1. Introduction

Interacting with nature helps restore attention[5], alleviate fatigue[1], and foster deep emotions like safety, happiness, and tranquility[6]. However, some groups have difficulty or infrequent contact with nature due to work conditions or physical limitations. Thus, using modern technology to simulate natural environments can be a viable alternative to real nature [3, 10, 12]. These simulated environments, known as Digital Nature, can reduce stress, enhance cognitive abilities, and positively affect physical and mental health.

Current research on digital nature has not focused on specific design elements, but studies on abstract qualities of nature[9] and environmental atmosphere characteristics[7] can serve as valuable references for designers. Our study aims to explore key elements in the design of immersive virtual nature (IVN) and their impact on user perception. By combining online surveys and experimental analysis, we identified elements from natural environments contributing to relaxation and recovery and validated these elements in IVN. We hope the critical element design strategies proposed in this study can help special groups achieve better relaxation and recovery effects when accessing digital nature. This paper focuses on the indoor natural scenes in the IVN experience based on head-mounted display (HMD) technology.

## 2. Immersive natural visual elements

### 2.1 Selection of Natural Videos

A recent empirical study[9] categorized eight relaxing qualities of digital nature into four themes: engaging, instinctive, ambient, and derivative. Through analysis of the interview content from this study, we found that the qualities of calming, immersing, reassuring, and recharging are included in multiple themes. Therefore, we used "calming," "immersing," "reassuring," and "recharging" as keywords, combined with "nature," to search for entries on public online sites pixabay.com and pexels.com, which provide copyright-free multimedia resources. We selected ten videos from each set of search results and edited them into continuous 15-second clips[13], resulting in 40 videos forming the video library for this study. Then, five designers were invited to rate these 40 videos on a 5-point scale. Finally, 10 videos were selected, as shown in Figure 1, where the scores represent the ratings in their respective categories.

* Corresponding author: chendengkai@nwpu.edu.cn

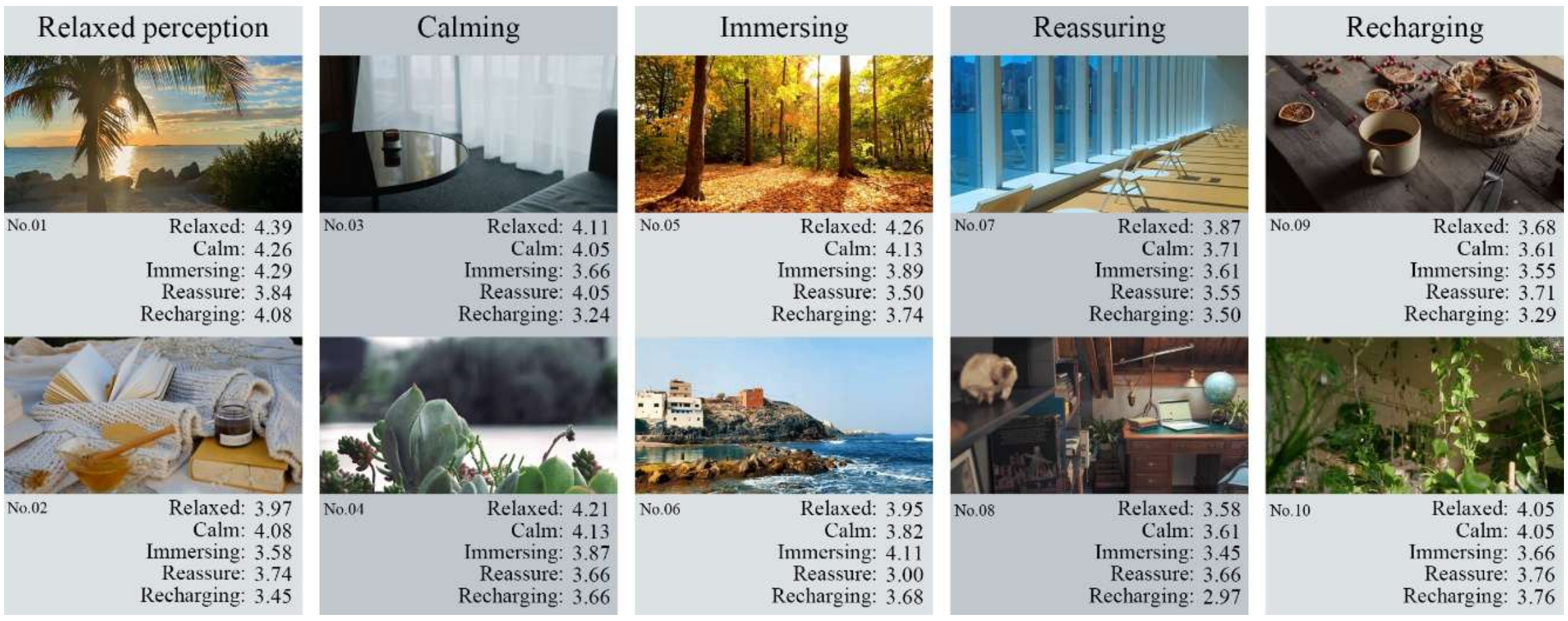

**Figure 1** Results of respondents' ratings of each question on the 10 videos.

## 2.2 Online Questionnaire

The questionnaire consists of three parts: participants' demographic information, the Nature Relatedness Scale[11], and the evaluation of the relaxing qualities of nature. In this study, using nature-relatedness helps to account for individual differences among respondents.

The questionnaire was created using the online survey design platform SurveyMonkey and was distributed via email to respondents in Asia, Europe, North America, and Oceania for one month. Similar to the questions used in the first round of designer screening, each video was accompanied by five Likert scale questions. Additionally, to extract important elements and prevent memory bias in post-interviews, we asked respondents to write down the elements that made them feel most relaxed after each video, with no word limit. To avoid interference from the environment on the respondents' emotional state during the questionnaire, we required participants to complete the survey in a quiet and solitary setting. Each video lasted 15 seconds, with no fast-forwarding or skipping allowed. Participants could only proceed to the next video after watching the current one and completing the rating. The entire questionnaire process took approximately 15 minutes.

We collected 50 complete responses for this survey. However, five questionnaires were discarded due to disorganized answers caused by incorrect question numbering. Additionally, based on the Nature Relatedness Scale results, six respondents scored below 3 points. Therefore, 39 valid questionnaires were retained.

## 2.3 Result Analysis

Figure 1 presents the relaxing ratings for all videos in the questionnaire and detailed scores for each relaxing quality. The author conducted a comprehensive analysis of the video ratings from the 39 valid respondents, categorizing the scores into two groups: overall relaxation perception evaluation and average relaxing quality evaluation. The average relaxing quality evaluation includes calm, immersing, reassuring, and recharging scores. Among all videos, calm (average: 3.94) and immersing (average: 3.77) received relatively high scores, while reassuring (average: 3.65) and recharging (average: 3.54) received relatively low scores in terms of the four relaxation qualities.

The online survey collected 719 terms describing relaxing elements in the videos. Figure 2 shows the word cloud based on word frequency. The survey responses included specific descriptions such as “warm-toned forest”, “minimalist desk”, “warm sunlight”, and “dynamic visual perception of rain on leaves”. We broke down these descriptions into concrete elements and perceptual descriptions. The most frequently mentioned concrete elements were grouped into four categories: “sunlight”, “water”, “plants”, and “furnishings” (including physical items, materials, textures, and colors).

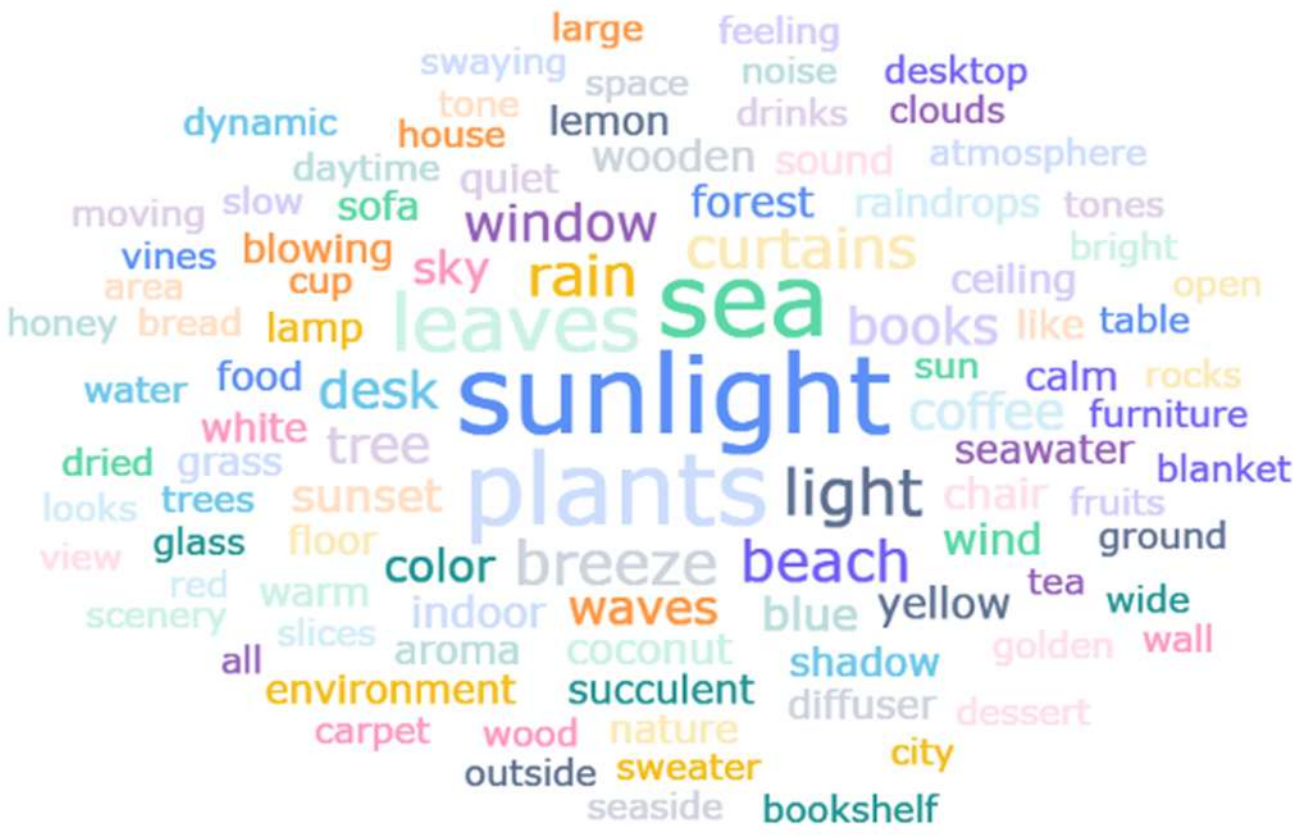

**Figure 2** Word Cloud Analysis Based on Word Frequency.

# 3. Immersive Elemental Validation Scenarios

## 3.1 Adaptive Design of Natural Elements

The online survey included video content featuring indoor and outdoor scenes, meaning that the scale and scope of the environments influenced the respondents' perceptions. The same elements can have different forms and scales in indoor versus outdoor settings. The biophilic design approach[4, 8] encompasses using natural elements through direct nature experience, indirect nature experience, and place-based experience, including both tangible physical elements and abstract emotional perceptions. Based on the biophilic design approach, this study maps the perceptual descriptions extracted from the survey to specific elements. It proposes corresponding design strategies for virtual environments, as shown in Figure 3. This paper primarily focuses on immersive virtual scenes designed for indoor environments. Our future work will expand the research to include indoor and outdoor environments.

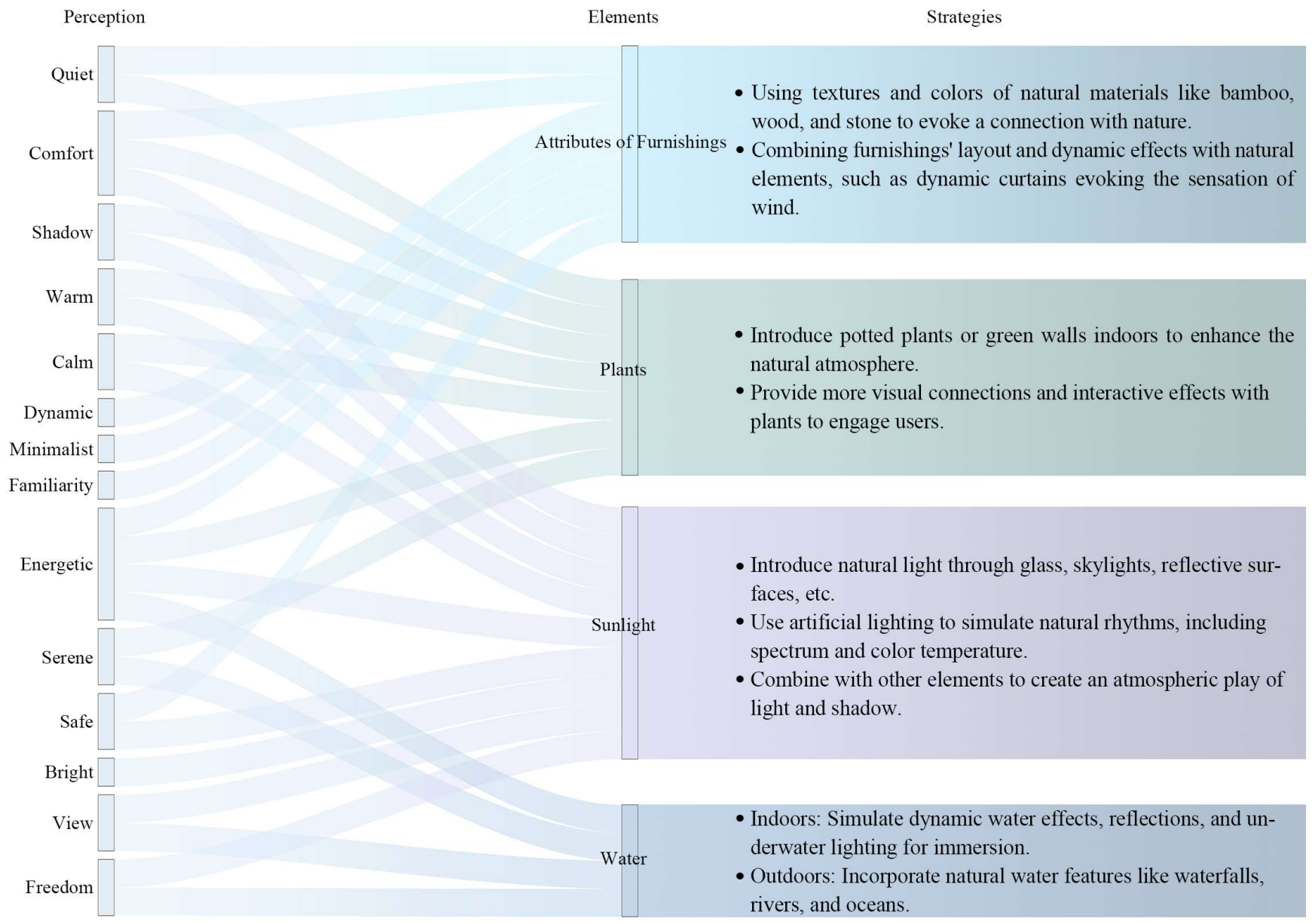

**Figure 3** Design elements and strategies for immersive virtual natural environments.

## 3.2 Stress-induced task design

We employed individual acute stress induction[15] to validate the effectiveness of immersive virtual environment elements for relaxation and recovery under laboratory conditions. Participants engaged in complex cognitive tasks with added sound and motivational punishment to induce stress. Based on common gaming and interactive behaviors and cognitive task paradigms, we selected three stress-inducing tasks: the stop-signal task[14] for response inhibition, the letter memory task[16] for memory updating, the trail-making test[2] for cognitive flexibility. All stress-inducing tasks were conducted within the immersive virtual environment to enhance the immersive virtual environment experience and maintain experimental continuity without repeatedly wearing the HMD. The task interface is shown in Figure 4.

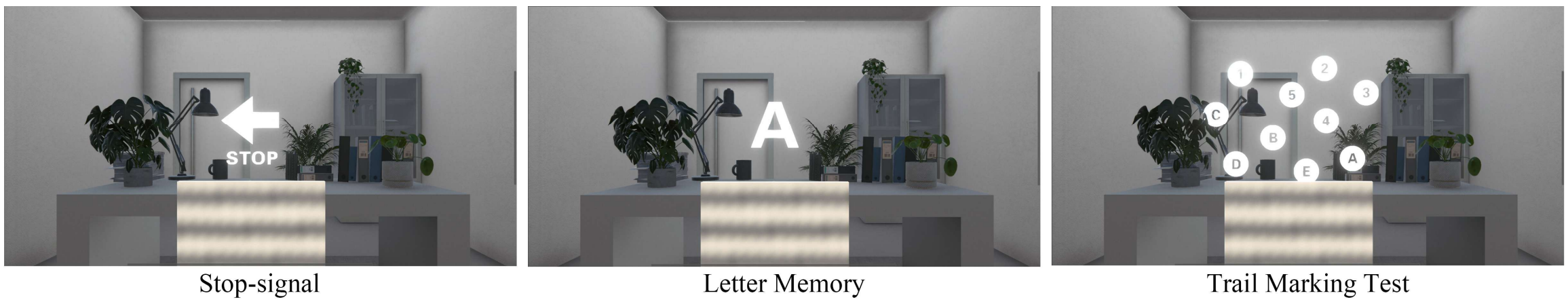

**Figure 4** Stress inducing task interface.

## 3.3 Immersive Indoor Virtual Natural Environment Experience

For an immersive indoor virtual environment experience, we created VR executables using Unity 3D (version 2022). We created a total of six scenes by controlling elemental variables.

We opted for common methods to represent the four categories of elements. For lighting, we incorporated closable windows in the model to switch between natural lighting scenarios, along with basic indoor lighting conducive to standard work requirements. We embedded water bodies for water elements and utilized volumetric lighting and scattering parameters to create the atmosphere generated by water. In terms of plants, we employed a combination of desktop and floor plants. The natural materials and biomimetic furniture enhanced the base scene, as shown in Figure 5.

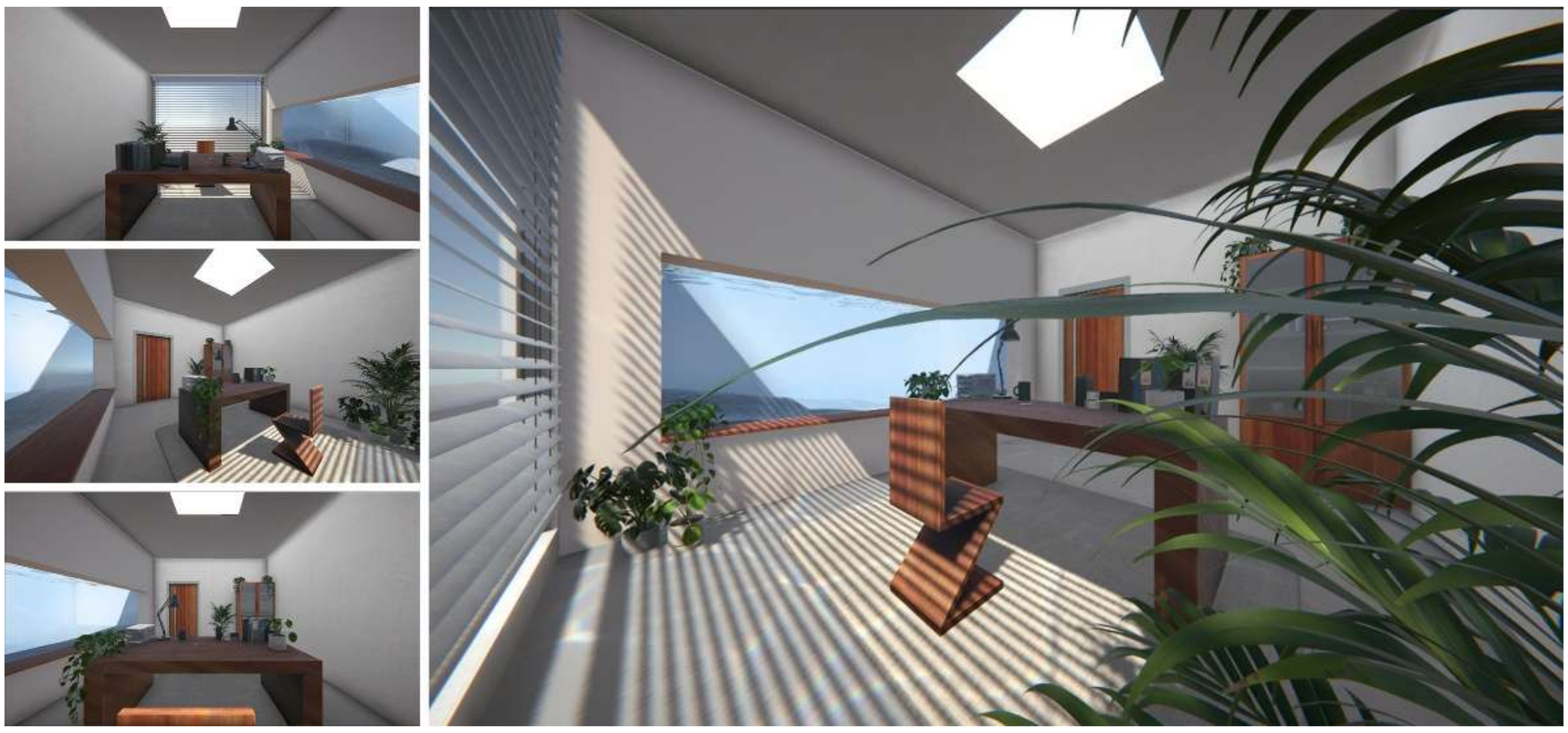

**Figure 5** IVN scenes used for experiments.

## 4. Next steps: validating the impact of immersive natural elements

### 4.1 Experimental Setup

The following validation phase will be conducted at the Key Laboratory of Industrial Design and Ergonomics, Ministry of Industry and Information Technology, Industrial Design Department, Northwestern Polytechnical University. We will recruit participants aged 18-35 for the experiment, screening them using the Nature Relatedness Scale. We will recruit 60 participants to ensure sufficient data for each experimental scenario.

### 4.2 Data Measurement Methods

This study will use wearable physiological measurement devices to continuously record participants' heart rate variability (HRV), heart rate (HR), and skin conductance level (SCL) in an immersive virtual environment, assessing acute stress responses and recovery. We will use the Tobii Pro VR eye tracker to collect participants' visual interest areas and gaze durations at a 120Hz sampling rate. Participants will complete the Self-Assessment Manikin (SAM) emotion scale before and after the experiment. After the stress induction and recovery scenarios, they will complete the short-form State-Trait Anxiety Inventory (STAI).

## 5. Conclusions

This study extracted concrete elements that characterize relaxing qualities through an online survey and proposed design strategies for immersive virtual natural scenes. We developed a series of immersive indoor nature virtual scene tools. Customized stress induction tasks stimulate participants to affect their emotions and stress, while natural elements in the immersive virtual scenes regulate their physiological and psychological states. Experimental results allow us to analyze these design elements' impact and optimal combination, providing reference strategies for human-computer interaction designers and interior designers.

## Acknowledgements

This study was funded by the Fundamental Research Funds for the Central Universities [G2023KY05105].

## References

[1] Archary, P. and Thatcher, A. 2022. Affective and cognitive restoration: comparing the restorative role of indoor plants and guided meditation. *Ergonomics*. 65, 7 (Jul. 2022), 933–942.

[2] Baykara, E., Kuhn, C., Linz, N., Troeger, J. and Karbach, J. 2022. Validation of a digital, tablet-based version of the Trail Making Test in the increment elta platform. *EUROPEAN JOURNAL OF NEUROSCIENCE*. 55, 2 (Jan. 2022), 461–467.

[3] Browning, M.H., Mimnaugh, K.J., Van Riper, C.J., Laurent, H.K. and LaValle, S.M. 2020. Can simulated nature support mental health? Comparing short, single-doses of 360-degree nature videos in virtual reality with the outdoors. *Frontiers in psychology*. 10, (2020), 2667.

[4] Browning, W.D. and Ryan, C.O. 2020. What is biophilia and what does it mean for buildings and spaces. *Nature inside: A Biophilic Design Guide. RIBA Publishing, pp. 1e5.*

[5] Hähn, N., Essah, E. and Blanusa, T. 2021. Biophilic design and office planting: a case study of effects on perceived health, well-being and performance metrics in the workplace. *Intelligent Buildings International*. 13, 4 (2021), 241–260.

[6] Herzog, T.R., Maguire, P. and Nebel, M.B. 2003. Assessing the restorative components of environments. *Journal of environmental psychology*. 23, 2 (2003), 159–170.

[7] van Houwelingen-Snippe, J., van Rompay, T.J.L. and Ben Allouch, S. 2020. Feeling Connected after Experiencing Digital Nature: A Survey Study. *International Journal of Environmental Research and Public Health*. 17, 18 (Jan. 2020), 6879.

[8] Kellert, S.R. 2018. *Nature by Design: The Practice of Biophilic Design*. Yale University Press.

[9] Kim, C.M., van Rompay, T.J.L., Louwers, G.L.M., Yoon, J. and Ludden, G.D.S. 2023. From a Morning Forest to a Sunset Beach: Understanding Visual Experiences and the Roles of Personal Characteristics for Designing Relaxing Digital Nature. *International Journal of Human–Computer Interaction*. 0, 0 (2023), 1–18.

[10] Kjellgren, A. and Buhrkall, H. 2010. A comparison of the restorative effect of a natural environment with that of a simulated natural environment. *Journal of environmental psychology*. 30, 4 (2010), 464–472.

[11] Nisbet, E.K. and Zelenski, J.M. 2013. The NR-6: a new brief measure of nature relatedness. *Frontiers in Psychology*. 4, (Nov. 2013).

[12] Yin, J., Zhu, S., MacNaughton, P., Allen, J.G. and Spengler, J.D. 2018. Physiological and cognitive performance of exposure to biophilic indoor environment. *Building and Environment*. 132, (Mar. 2018), 255–262.

[13] Zhu, C., Xu, X., Zhang, W., Chen, J. and Evans, R. 2020. How Health Communication via Tik Tok Makes a Difference: A Content Analysis of Tik Tok Accounts Run by Chinese Provincial Health Committees. *International Journal of Environmental Research and Public Health*. 17, 1 (Jan. 2020), 192.

[14] Fang, J., Zhu, Y., Zhao, W., Zhang, B. and Wang, X. 2013. Stop Signal Task and the Related Models of Response Inhibition. Chinese Journal of Clinical Psychology. 21, 5 (2013), 743-746+750.

[15] Duan, H., Wang, X., Wang, B., WANG, T., Zhang, X., Wang, Z. and Hu, W.2017. Acute stress: Induction, measurement and effect analysis. Advances in Psychological Science. 25, 10 (2017), 1780–1790.

[16] Zhao, X. and Zhou, R. 2011. Sub-function Evaluation Method of Central Executive System in Working Memory. Chinese Journal of Clinical Psychology. 19, 6 (2011), 748–752